\documentclass[letters,traditabstract,longauth]{aa} 
\usepackage{graphicx}
\usepackage{txfonts}

\newcommand{\td}{$\mathrm{T_d}$}
\newcommand{\bt}{\begin{tabular}}
\newcommand{\et}{\end{tabular}}
\begin{document}
\title{{\it\bf Herschel}-ATLAS: Dust temperature and redshift distribution of SPIRE and PACS detected sources using submillimetre colours\thanks{{\it Herschel} is an ESA space observatory with science instruments provided by European-led Principal Investigator consortia and with important participation from NASA.}}

\author{
A. Amblard\inst{1}
\and
A. Cooray \inst{1}
\and
P. Serra \inst{1}
\and
P. Temi\inst{2}
\and
E. Barton\inst{1}
\and
M. Negrello\inst{3}
\and
R. Auld\inst{4}
\and
M. Baes\inst{5}
\and
I.K. Baldry\inst{6}
\and
S. Bamford\inst{7}
\and
A. Blain\inst{8}
\and
J. Bock\inst{9}
\and
D. Bonfield\inst{10}
\and
D. Burgarella\inst{11}
\and
S. Buttiglione\inst{12}
\and
E. Cameron\inst{13}
\and
A. Cava\inst{14}
\and
D. Clements\inst{15}
\and
S. Croom\inst{16}
\and
A. Dariush\inst{4}
\and
G. de Zotti\inst{12,21}
\and
S. Driver\inst{17}
\and
J. Dunlop\inst{18}
\and
L. Dunne\inst{7}
\and
S. Dye\inst{4}
\and
S. Eales\inst{4}
\and
D. Frayer\inst{19}
\and
J. Fritz\inst{5}
\and
Jonathan P. Gardner\inst{20}
\and
J. Gonzalez-Nuevo\inst{21}
\and
D. Herranz\inst{22}
\and
D. Hill\inst{17}
\and
A. Hopkins\inst{23}
\and
D. H. Hughes\inst{24}
\and
E. Ibar\inst{25}
\and
R.J. Ivison\inst{25}
\and
M. Jarvis\inst{10}
\and
D.H. Jones\inst{23}
\and
L. Kelvin\inst{17}
\and
G. Lagache\inst{26}
\and
L. Leeuw\inst{2}
\and
J. Liske\inst{27}
\and
M. Lopez-Caniego\inst{22}
\and
J. Loveday\inst{28}
\and
S. Maddox\inst{7}
\and
M. Micha{\l}owski\inst{18}
\and
P. Norberg\inst{18}
\and
H. Parkinson\inst{18}
\and
J.A. Peacock\inst{18}
\and
C. Pearson\inst{29,30}
\and
E. Pascale\inst{4}
\and
M. Pohlen\inst{4}
\and
C. Popescu\inst{31}
\and
M. Prescott\inst{6}
\and
A. Robotham\inst{17}
\and
E. Rigby\inst{7}
\and
G. Rodighiero\inst{32}
\and
S. Samui\inst{21}
\and
A. Sansom\inst{31}
\and
D. Scott\inst{33}
\and
S. Serjeant\inst{3}
\and
R. Sharp\inst{23}
\and
B. Sibthorpe\inst{25}
\and
D.J.B. Smith\inst{7}
\and
M.A. Thompson\inst{10}
\and
R. Tuffs\inst{34}
\and
I. Valtchanov\inst{35}
\and
E. Van Kampen\inst{27}
\and
P. Van der Werf\inst{36}
\and
A. Verma\inst{37}
\and
J. Vieira\inst{8}
\and
C. Vlahakis\inst{36}
}

\institute{
Dept. of Physics \& Astronomy, University of California, Irvine, CA 92697, USA
\and
Astrophysics Branch, NASA Ames Research Center, Mail Stop 245-6, Moffett Field, CA 94035, USA
\and
Dept. of Physics and Astronomy, The Open University, Milton Keynes, MK7 6AA, UK
\and
School of Physics and Astronomy, Cardiff University, The Parade, Cardiff, CF24 3AA, UK
\and
Sterrenkundig Observatorium, Universiteit Gent, Krijgslaan 281 S9,B-9000 Gent, Belgium
\and
Astrophysics Research Institute, Liverpool John Moores University,12 Quays House, Egerton Wharf, Birkenhead, CH41 1LD, UK
\and
School of Physics and Astronomy, University of Nottingham, University Park, Nottingham NG7 2RD, UK
\and
Caltech, 249-17, Pasadena, CA 91125, USA
\and
Jet Propulsion Laboratory, Pasadena, CA 91109, United States; Department of Astronomy, California Institute of Technology, Pasadena, CA 91125, USA
\and
Centre for Astrophysics Research, Science and Technology Research Centre, University of Hertfordshire, Herts AL10 9AB, UK
\and
Laboratoire d'Astrophysique de Marseille, UMR6110 CNRS, 38 rue F.Joliot-Curie, F-13388 Marseille France
\and
INAF - Osservatorio Astronomico di Padova,  Vicolo Osservatorio 5, I-35122 Padova, Italy
\and
ETH Zurich, Insitute for Astronomy, HIT J12.3, CH-8093 Zurich, Switzerland
\and
Instituto de Astrof\'{i}sica de Canarias, C/V\'{i}a L\'{a}ctea s/n, E-38200 La Laguna, Spain\\
Departamento de Astrof{\'\i}sica, Universidad de La Laguna (ULL), E-38205 La Laguna, Tenerife, Spain
\and
Astrophysics Group, Imperial College, Blackett Laboratory, Prince Consort Road, London SW7 2AZ, UK
\and
Sydney Institute for Astronomy, School of Physics, University of Sydney, NSW 2006, Australia
\and
SUPA, School of Physicsand Astronomy, University of St. Andrews, North Haugh, St. Andrews, KY169SS, UK
\and
Scottish Universities Physics Alliance, Institute for Astronomy, University of Edinburgh, Edinburgh, EH9 3HJ, UK
\and
Infrared Processing and Analysis Center, California Institute of Technology, 770 South Wilson Av, Pasadena, CA 91125, USA
\and
NASA Goddard Space Flight Center, Greenbelt, MD USA 20771
\and
Scuola Internazionale Superiore di Studi Avanzati, via Beirut 2-4,34151 Triest, Italy
\and
Instituto de F\'isica de Cantabria (CSIC-UC), Santander, 39005, Spain
\and
Anglo-Australian Observatory, PO Box 296, Epping, NSW 1710, Australia
\and
Instituto Nacional de Astrof\'{i}sica, \'{O}ptica y Electr\'{o}nica (INAOE),Luis Enrique Erro No.1, Tonantzintla, Puebla, C.P. 72840, Mexico
\and
UK Astronomy Technology Center, Royal Observatory Edinburgh, Edinburgh, EH9 3HJ, UK
\and
Institut d'Astrophysique Spatiale (IAS), B\^{a}timent 121, F-91405 Orsay, France; and Universit\'{e} Paris-Sud 11 and CNRS (UMR 8617), France
\and
European Southern Observatory, Karl-Schwarzschild-Strasse 2 D-85748, Garching bei Munchen, Germany
\and
Astronomy Centre, University of Sussex, Falmer, Brighton, BN1 9QH, UK
\and
Space Science \& Technology Department CCLRC RAL, Rutherford Appleton Laboratory Oxon, OX11 0QX, UK
\and
Institute for Space Imaging Science, University of Lethbridge, Lethbridge, Alberta CANADA, T1K 3M4
\and
Jeremiah Horrocks Institute, University of Central Lancashire, Preston, Lancs PR1 2HE, UK
\and
University of Padova, Department of Astronomy, Vicolo Osservatorio 3, I-35122 Padova, Italy
\and
Dept. of Physics \& Astronomy, University of British Columbia, 6224 Agricultural Road, Vancouver, B.C. V6T 1Z1, Canada
\and
Max Planck Institute for Nuclear Astrophysics (MPIK), Saupfercheckweg 1,69117 Heidelberg, Germany
\and
Herschel Science Centre, ESAC, ESA, PO Box 78, Villanueva de la Ca\~nada, 28691 Madrid, Spain
\and
Sterrewacht Leiden, Leiden University, PO Box 9513, 2300 RA Leiden, The Netherlands
\and
Oxford Astrophysics, Denys Wilkinson Building, University of Oxford, Keble Road, Oxford, OX1 3RH, UK
}

  \abstract
{
 We present colour-colour diagrams of detected sources in the {\it Herschel}-ATLAS Science Demonstration Field
from $100$ to $500\,\mu$m using both PACS and SPIRE. 
We fit isothermal modified black bodies to the spectral energy distribution (SED) to extract
the dust temperature of sources with counterparts in Galaxy And Mass Assembly (GAMA) or SDSS surveys with either a spectroscopic or a photometric redshift.
For a subsample of 330 sources detected in at least three FIR bands with a significance greater than 3$\sigma$, we find an average dust temperature of $(28 \pm 8)$K.
For sources with no known redshift, we populate the colour-colour diagram with a large number of SEDs generated with a broad range of dust temperatures and emissivity parameters,
and compare to colours of observed sources to establish the redshift distribution of this sample. 
For another subsample of 1686 sources with fluxes above 35 mJy at $350\,\mu$m and detected at 250 and $500\,\mu$m with a significance greater than 3$\sigma$, we find 
an average redshift of $2.2 \pm 0.6$.
}

\keywords{submillimeter: galaxies --- Galaxies: evolution --- Galaxies: high-redshift}

\titlerunning{Colour-colour diagrams of {\it Herschel}-ATLAS sources}

\maketitle
%

\section{Introduction}

The total intensity of the extragalactic background light at far-infrared (FIR) wavelengths has been measured 
with absolute photometry (Puget et al. 1996; Fixsen et al.  1998; Dwek et al. 1998). Deep surveys are 
now starting to resolve the cosmic FIR background into discrete sources 
with the fraction resolved varying with wavelength
(see reviews in Blain et al. 2002; Hauser $\&$ Dwek 2001;  Lagache et al. 2005, Dye et al. 2009).
Despite some successes, there are still large uncertainties about the nature and evolution of the 
submillimetre (submm) galaxy population, including their 
redshift distribution.

The {\it Herschel}-Astrophysical Terahertz Large Area Survey (H-ATLAS)  is an open-time key program of the 
{\it Herschel} Observatory 
(Pilbratt et al. 2010) that will survey roughly 550 deg$^2$ of sky over 600 hours of
observations (Eales et al. 2010). To cover the widest possible area, H-ATLAS  uses the maximum 
possible scan rate for the telescope at 60 arcsec\,sec$^{-1}$ 
in the parallel mode of
PACS (Poglitsch et al. 2010) and SPIRE  (Griffin et al. 2010). The observations cover five 
photometric bands with $100$ and $160\,\mu$m data from  PACS, and  
$250$, $350$, and $500\,\mu$m data from SPIRE. The H-ATLAS fields include 1 field close to the 
northern Galactic pole (150 deg$^2$), 3 fields 
(each 36 deg$^2$) coinciding with the GAMA redshift survey fields (Driver et al. 2009), and 2 fields 
(total area of 250 deg$^2$) near the southern Galactic pole. 

During the Science Demonstration Phase (SDP), H-ATLAS observed 14.4 deg$^2$ in the GAMA-9 hour
 field near the ecliptic plane to a 5$\sigma$ depths of 35-90 mJy.  In this {\it Letter} we  
present the colour-colour diagrams derived from the
 five bands from $100$ to $500\,\mu$m
for the H-ATLAS sources that are cross-identified in at least 3 of the SPIRE and PACS bands.
  We then discuss the
dust temperature and redshift distributions based on
certain assumptions about the spectral energy distribution (SED) and subject to selection effects
 associated with the sample selection.

\begin{figure}[h!]
  \centering
  \includegraphics[width=8cm]{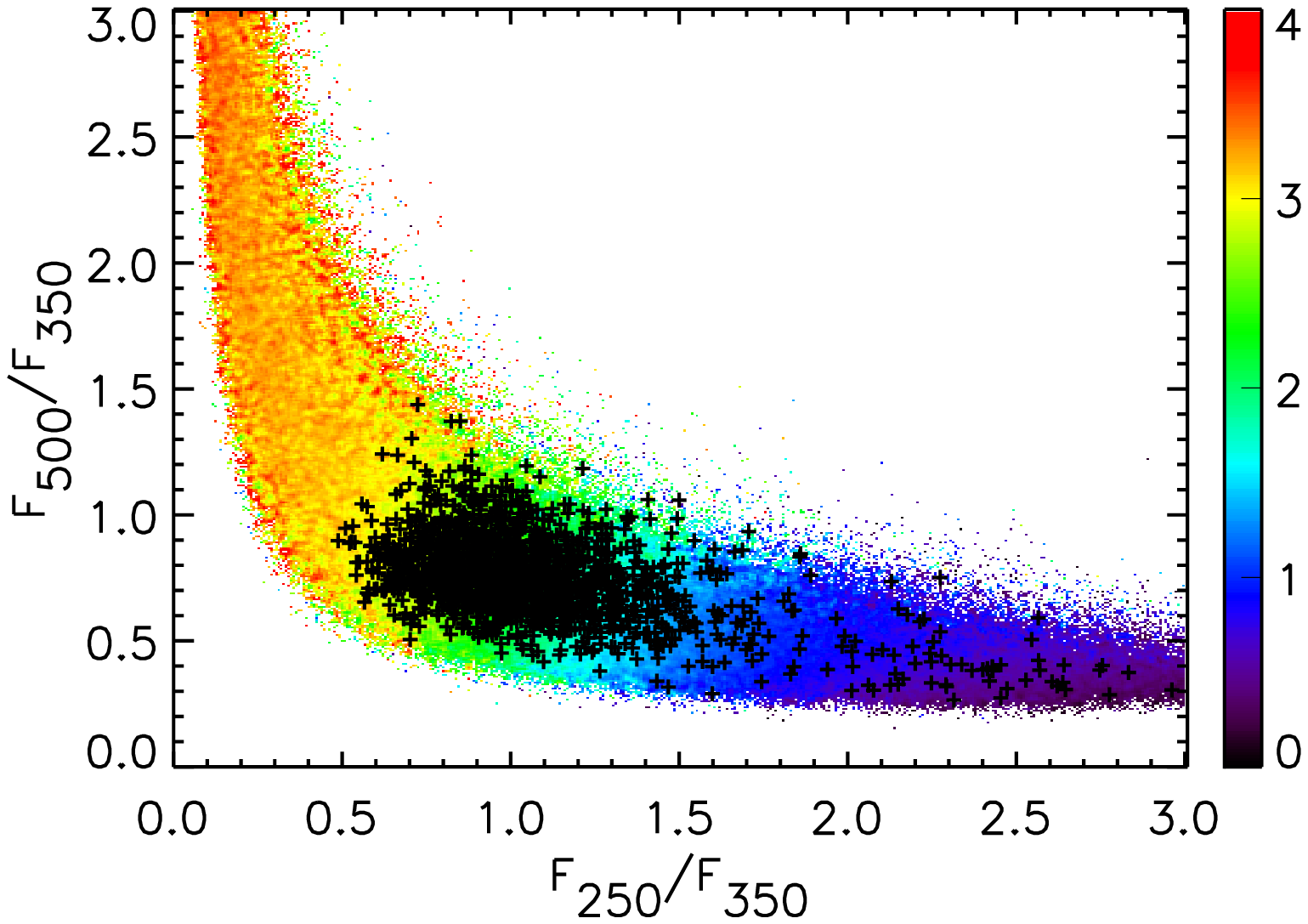}
  \includegraphics[width=8cm]{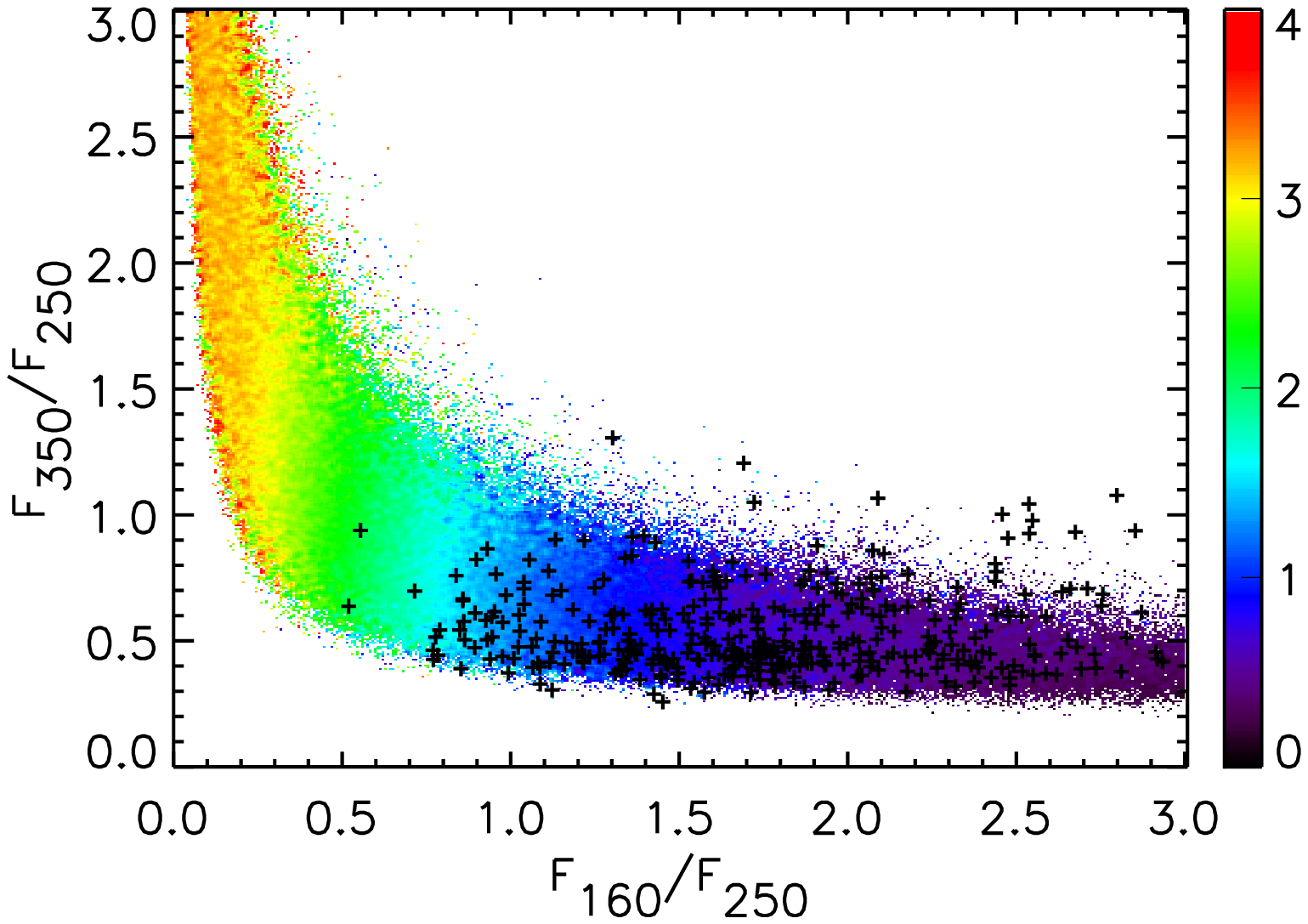}
  \includegraphics[width=8cm]{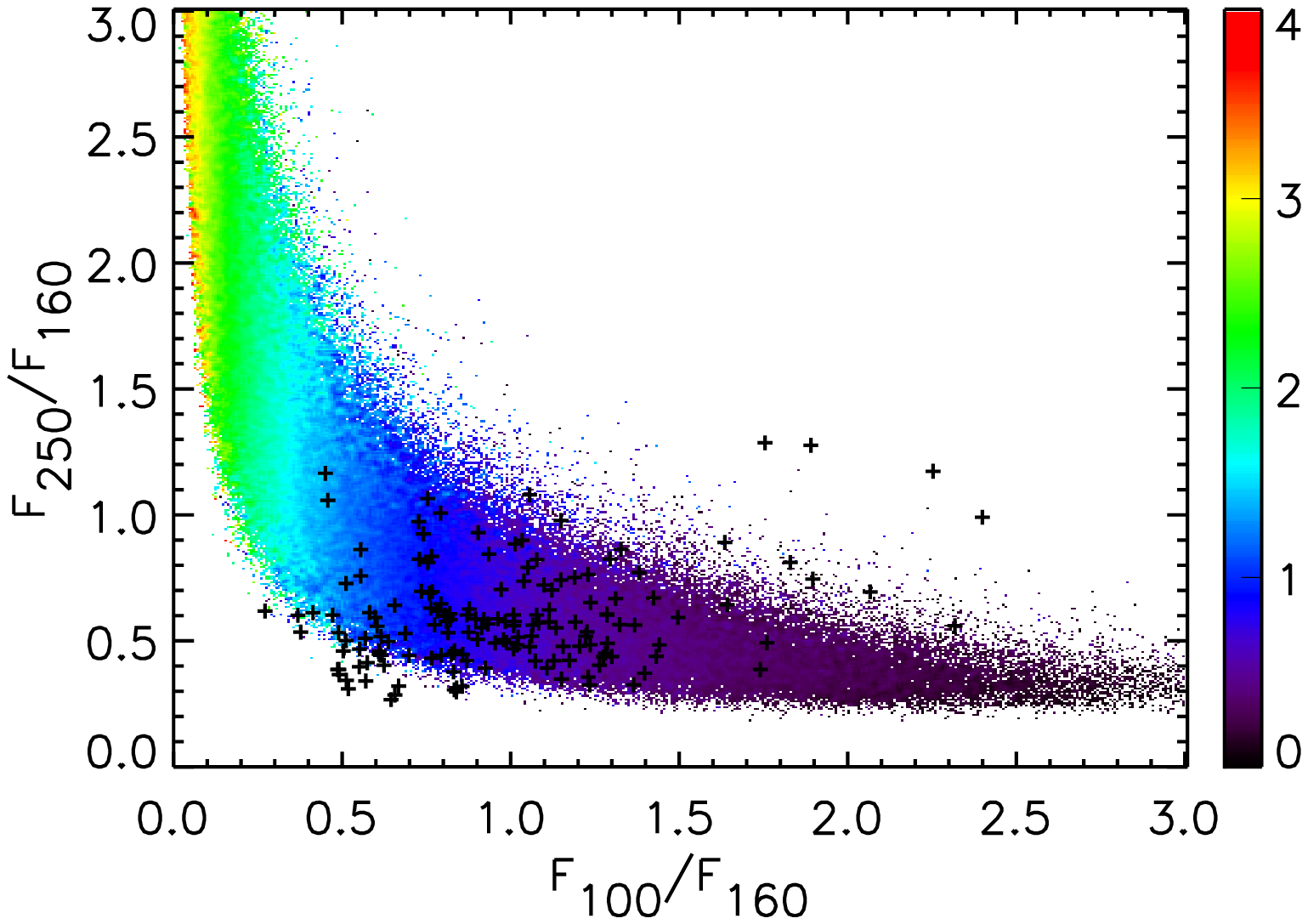}
  \caption{Colour-colour diagram of SPIRE and PACS sources. 
   The coloured background indicates the average redshift in these colour-colour
  spaces of our 10$^6$ randomly generated model SEDs. In (a) points plotted represent
  a sample of 1686 galaxies detected in all three SPIRE bands with a significance greater than
 5 $\sigma$ at $350\,\mu$m ($S>35$mJy)
and greater than 3$\sigma$ at $250$ and $500\,\mu$m (corresponding to flux limits of roughly 21 and 
27 mJy, respectively).
In (b) we show a sample of 402 galaxies detected at $160$, $250$, and $350\,\mu$m 
and in (c) we show a sample of 158 galaxies detected at $100$, $160$, and $250\,\mu$m. As in (a), 
in both (b) and (c) we also select sources
by imposing a 5$\sigma$ cut in the central wavelength of the colour plot and greater than 3$\sigma$ at other wavelengths.
}
  \label{colcol}
\end{figure}

\begin{figure}
  \centering
  \includegraphics[width=9cm]{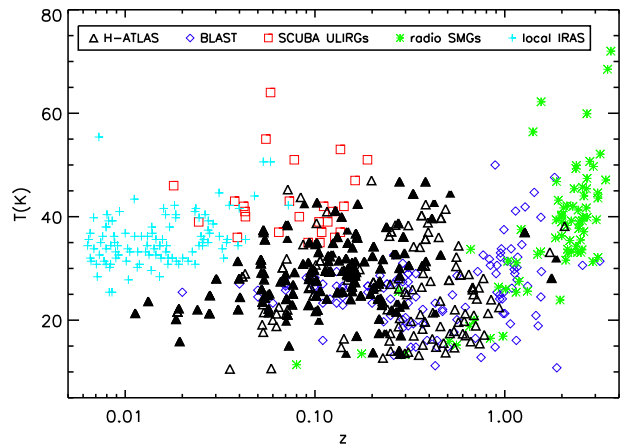}
  \caption{Dust temperature as a function of redshift for 330 H-ATLAS sources reliably identified 
with GAMA and SDSS DR-7 galaxies having a known spectroscopic (black filled triangle) or 
a photometric (black empty triangle) redshift.
The selected sources are $>5\sigma$ detections in at least one band of  either PACS or SPIRE with
 $3 \sigma$ flux measurements in at least 2 more bands. The isothermal 
spectral fits assume $\beta=1.5$. For other data plotted, see Section~3.}
\label{Tvsz}
\end{figure}

\begin{figure}
\centering
\includegraphics[width=9cm]{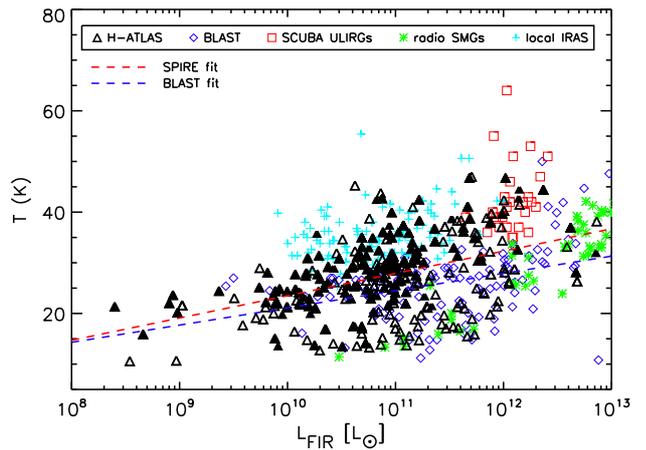}
\caption{Dust temperature as a function of the FIR luminosity for 330 H-ATLAS sources 
reliably identified with GAMA and SDSS DR-7 galaxies with a known spectroscopic (black filled triangle)
 or a photometric (black empty triangle) redshift. The FIR luminosity is obtained by 
fitting the PACS and SPIRE measurements with a modified black body (parameters are the dust
 temperature and the luminosity density, $\beta$ is fixed to 1.5) and this model SED is integrated
 between $8$ and $1100\,\mu$m. We also indicate BLAST and H-ATLAS best-fit $T_d-\log (L_{FIR})$ relation
with  a dashed blue line (from Dye et al. 2009) and a dashed red line, respectively.
}
\label{TvsL}
\end{figure}

\section{Colour-colour diagrams}

We use source catalogues generated for the H-ATLAS consortium by Rigby et al. (2010, in prep.),
 derived from
 SPIRE and PACS maps presented in Pascale et al. (2010, in prep.) and Ibar et al. (2010, in prep.), respectively.
The source catalogues are supplemented with
cross-identification information from the GAMA survey (Driver et al. 2009) and SDSS DR-7 (Abazajian et al. 2009)
 as described in Smith et al. (2010, in prep.). For the sources for which spectroscopic redshifts from GAMA or SDSS are
not available, we use photometric redshifts  generated using the ANNz
neural network code (Collister \& Lahav 2004), trained with photometry
from SDSS and UKIDSS LAS (Lawrence et al. 2007), and spectra from the
GAMA spectroscopic survey (Driver et al 2009), DEEP2 (Davis et al. 2007), zCOSMOS (Lilly et al. 2007), and the 2SLAQ-LRG (Cannon et al. 2006) survey.

In Fig.~1, we show the colour-colour plots of the H-ATLAS sources. We divide these plots 
in terms of colours  based on either SPIRE only  or SPIRE and PACS data. Our flux selection results in 
selecting 1686, 402, and 158 sources, from top to bottom of Fig.~1. 
For reference, the total ATLAS catalogue for this field contains $\sim$ 6600 sources (Rigby  et al. 2010, in prep.).

The colour-colour plots are filled with 10$^6$ black-body spectra 
at a single dust temperature, $T_{\rm d}$, modified by a frequency-dependent 
emissivity function $\epsilon_{\nu}\propto \nu^{\beta}$, where the flux density $f_\nu$ is 
\begin{equation}
f_{\nu}=\epsilon_{\nu}B_{\nu}\propto\nu^{3+\beta}/[\exp\Big(\frac{h\nu}{kT_{\rm d}}\Big)-1].
\end{equation}
In generating these models, we consider uniform ranges of dust temperature from 10 to 60 K, 
emissivity parameter $\beta$ from 0 to 2,
and redshift from 0 to 5. The choice of $0 < \beta <2$ compared to $1 < \beta < 2$ makes a minor difference
since we also broaden the SED tracks in the colour diagram by adding an extra 
Gaussian standard deviation of 10\% to the fluxes used to compute the model colour. 
This scatter accounts for the broadening of data in the colour-colour plane caused by flux uncertainties. 
As shown in Fig.~1, we find that the colour diagram of sources with fluxes only from SPIRE are well 
within the limits defined by the models we have considered.
When we examine PACS colours, we find that some points lie outside the same set of tracks as used for
 the SPIRE-only colour diagram. While some of these outlier points may 
be cause by either the fractionally larger flux errors of PACS or contamination from
a neighboring source, it is possible that some of these sources
are not accurately described by our simple isothermal SED model, requiring for instance a second dust component
(Dunne $\&$ Eales 2001) or a more complex SED model. 

When fitting a simple modified black-body model to the data, we must keep in mind that there is a 
partial degeneracy between $\beta$ and $T_{\rm d}$ and, more importantly, 
a perfect degeneracy between $T_{\rm d}$ and $z$. The peak of the SED is 
determined by the $\nu/T_{\rm d}$ term in the exponential,  so that a measurement of the colours alone 
constrains only the ratio $(1+z)/T_{\rm d}$. However, assuming reasonable priors on the free parameters of 
the SED model (in our case, $\beta$ and $T_{\rm d}$) it is still possible to estimate
 a qualitative redshift distribution for our sample of sources. Alternatively, if secure redshifts are known from 
optical cross-identifications, we can determine the
dust temperatures.

\begin{figure}
  \centering
  \includegraphics[width=8cm]{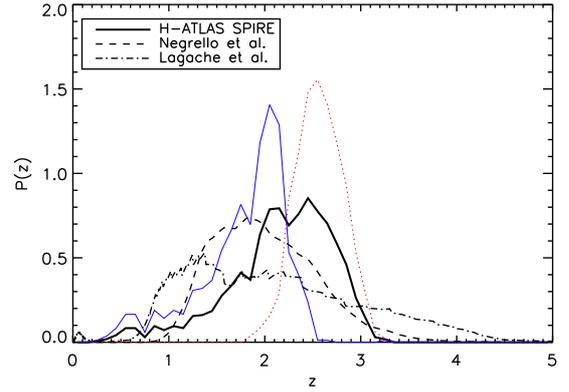}
  \caption{Normalized redshift distribution (black thick solid line) of SPIRE sources 
  in the colour-colour diagram of Fig.~1(a), selected to satisfy the following criteria:
  $S_{\rm 350}>35$mJy and at least 3$\sigma$ detections at 250 and $500\,\mu$m.
  The dashed line and dotted-dash line represent Negrello et al. (2007) and Lagache et al. (2004) models for
  the same flux selection. Both model curves have similar but not identical distributions, although Lagache
  model predicts a larger number of high-redshift galaxies. In addition, we show $N(z)$ for the same subsample using  a colour-cut of $S_{500}$/$S_{250} <$ 0.75 (blue thin solid line), 
  and $>0.75$ (red thin dotted line). The distribution shown here is for 1686 sources, a subset of about 6600 H-ATLAS sources in the SDP field.
}
  \label{dndz}
\end{figure}

\section{Dust temperature distribution}

We consider the H-ATLAS source sample with detections at $3\sigma$ in at least two bands and 
at $5\sigma$ in one band
of either PACS and SPIRE that have been robustly (reliability parameter R$_{\rm LR}>0.9$, Smith et al. 2010, in prep.) 
identified with GAMA or SDSS DR-7 galaxies. We also 
require that there is a known spectroscopic redshift
from either GAMA or SDSS, or from the photometric redshift catalog (with $(1+z)/\sigma_z>5$ and $z/\sigma_z>1$) 
that was generated for this field and cross-identified with  ATLAS sources (Smith et al. 2010, in prep.).
We select 330 sources, which correspond to the low redshift subsample of the galaxies selected in
 the colour-colour diagrams. For each galaxy, we perform a single temperature fit from the above equation 
assuming $\beta=1.5$.
We found that an isothermal SED model generally is a good fit to the 330 galaxies. 
Fitting a two-component SED model does not, on average, seem
to provide a closer fit.  The relatively good fit provided by the isothermal SED model
is probably due to most of our sample consisting of low redshift galaxies and therefore that we do not probe
much of the Wien part of the galaxy SED. 

In Fig.~2, we summarize our results and compare  
the H-ATLAS dust temperatures with dust temperatures in the literature for a variety of sub-mm 
bright galaxies.
These samples are: (i) the sources in BLAST detected above 5$\sigma$ in at least one of the BLAST bands 
with either a COMBO-17 (Wolf et al. 2004) or a SWIRE photometric redshift (Rowan-Robinson et al. 2008) and {\it Spitzer}-MIPS $70$ and $160\,\mu$m fluxes (Dye 
et al. 2009; $\beta=1.5$ fixed); (ii)
local ULIRGS observed with SCUBA at $450$ and $850\,\mu$m and complemented with IRAS $60$ and 
$100\,\mu$m fluxes (Clements et al. 2010; $\beta$ varied); (iii)
SCUBA sub-mm galaxies detected at better than 3$\sigma$ at $850\,\mu$m and having redshifts determined
from Keck-I spectroscopy (Chapman et al. 2005; $\beta=1.5$ fixed); and (iv) local IRAS-selected galaxies with 60
 and $100\,\mu$m fluxes complemented with SCUBA $850\,\mu$m (Dunne et al. 2000;
$\beta$ varied).

In Table~1, we list average dust temperatures as a function of redshift for several bins in redshift
 for both H-ATLAS only and all of the combined sub-mm galaxy samples, including H-ATLAS,
plotted in Fig.~2.  We do not find any evolution in the H-ATLAS dust temperature with redshift, though 
some evolution is inferred by BLAST measurements 
(Dye et al. 2009; Pascale et al. 2009), where sources at higher redshift had higher temperature in agreement
with the radio-identified submillimeter-selected galaxies (SMGs) found with SCUBA (Chapman et al. 2005; Ivison et al. 2010; Kovacs et al. 2006; Coppin et al. 2008).
While we find remarkable consistency between the average dust temperatures for our H-ATLAS sample as a
 function of redshift, these average values are not necessarily in agreement with other
sub-mm galaxy subsamples in the literature, mostly due to selection effects.
For example, the SCUBA ULIRGS have an average temperature of $(43 \pm 7)$K.
These sources were selected using IRAS $60\,\mu$m data and IRAS selected
sources are known to be biased towards higher temperatures.
Optically-selected low-redshift sub-mm  galaxies are known to have colder temperatures
consistent with our findings (e.g., Willmer et al. 2009; Vlahakis et al. 2005).
The large expected sample of sources from the 550 deg.$^2$ of H-ATLAS
will enable more detailed studies in the future, 
without the biases associated with selections of various sub-samples, including our own.
While we show results here with $\beta=1.5$ fixed, when fitting for $\beta$ and $T_{\rm d}$
 we found $\beta = 1.4 \pm 0.1$, consistent with $\beta=1.3$ 
found in Dunne et al. (2000).

In Fig.~3, we plot the dust temperature versus FIR luminosity by integrating model SEDs between 8 and 1100 $\mu$m
for the H-ATLAS subsample. Luminosities for other samples are from the literature.
Fitting for a relation of the form $T_{\rm d}={T_{\rm 0}}+\alpha\log(L_{FIR}/L_{\odot})$, we find ${T_{\rm 0}}=-20.5$K 
and $\alpha=4.4$ (see, Fig.~4). The value of $\log(L_{FIR}/L_{\odot})$ is on average $10.9\pm0.8$ for our sample.
This relation is consistent with the BLAST data (Dye et al. 2009).

\begin{table}
\caption{Average dust temperatures as a function of redshift for the 330 H-ATLAS galaxies
(Col. 2) and for all the data (Col. 4) presented in Fig.~2 and 3 (including H-ATLAS).}
\begin{center}
{\tiny
\begin{tabular}{ccccc}
\hline
$z$-range &\bt{c}H-ATLAS\\N$_{\rm srcs}$\et & \td  &\bt{c}all data\\ N$_{\rm srcs}$\et & all data \td \\
\hline
All z & 330 & $28\pm 8$ & 657 & $30\pm 9$\\
$0<z<0.1$ & 106 & $27\pm 8$  &  235 & $32 \pm 9$ \\ 
$0.1<z<0.5$ & 186  &  $29\pm 8$  &  260 & $28 \pm 8$ \\
$0.5<z<1$ & 33  &  $23 \pm 5$  &  67 & $24 \pm 7$ \\
$z>1$ & 5  &  $35\pm 4$  &  95 & $37 \pm 10$ \\
\hline
 \end{tabular}
}
 \label{table:1}
 \end{center}
 \end{table}

\section{Redshift distribution}

Since most of our sources lack redshifts, we consider another example application of 
our colour diagram and  infer the statistical redshift distribution $N(z)$ for the  source samples
plotted in Fig.~1. We do this by first gridding the colour-colour plane along the SED tracks into redshift bins.
 We then convert the number of sources within a grid region of colours to a binned redshift distribution 
(Hughes et al. 2002). This method is equivalent to extracting the redshift probability distribution function
 for the whole sample if we had simply fitted SEDs to individual fluxes and taken the
 sum of the redshift probabilities of each source. 
While the redshift for an individual galaxy is largely uncertain, and sensitive to the SEDs used,
 the statistical redshift distribution we extract should be a reasonable estimate of the true  
distribution of the source sample.

Figure~4 shows the redshift distribution for SPIRE sources detected in all 3 bands (Fig.~1(a)).  
For the sample of 1686 sources with flux densities above 35 mJy at $350\,\mu$m and above  
3$\sigma$ at $250$ and $500\,\mu$m, we find the average redshift to be $2.2 \pm 0.6$. This is consistent with the 
average redshift of $2.4 \pm 0.4$ for the radio-identified sub-mm galaxies with SCUBA at  $850\,\mu$m  followed up spectroscopically at Keck
 (Chapman et al. 2005), the interquartile redshift range of 1.8 to 3.1 for sources in SCUBA Half Degree Extragalactic
 Survey determined photometrically  (Aretxaga et al. 2007),
and a median redshift of greater than 1 for a subsample of 250 $\mu$m sources as faint as 35 mJy (Dunlop et al. 2009).
Our distribution, however, has an average redshift that is higher than that 
of a larger sample of  BLAST sources (Dye et al. 2009). This difference is not unexpected given the differences in the sample selection.   We emphasize again that the distribution shown in Fig.~4
is for a subset of about 25\% of the whole catalog 
with specific selection criteria that tend to favour the high redshift-end of the distribution.
Based on cross-identifications with SDSS and GAMA, we find that 30\% of the H-ATLAS sources are at $z < 0.5$ in the SDP field (Smith et al. 2010, in prep.). 

In addition to the total sample, we also make a colour-cut with $S_{500}/S_{250} > 0.75$ and $<0.75$,
 in addition to the flux cut of  35 mJy at 350$\mu$m.
By imposing this colour-cut, we distinguish a high redshift population with redder colours from a
 low redshift population. We find average redshifts of $2.6 \pm 0.3$ and $1.8 \pm 0.4$
for sources with  $S_{500}/S_{250} > 0.75$ and $<0.75$, respectively.
While  we do not show the redshift distributions separately for the samples in Figs.~1(b) and (c),
the average redshift  in these cases is  $z=0.8 \pm 0.3$ for sources detected in $160$, 
$250$, and $350\,\mu$m with a flux cut of 35 mJy at $250\,\mu$m,
and $z=0.7 \pm 0.3$ for sources detected in $100$, $160$, and $250\,\mu$m with a flux cut of $\simeq$ 100 mJy at
 $160\,\mu$m. As is clear from Fig.~1, we find that lower redshift sources are more likely to be detected
at shorter wavelengths, while bright sources seen in longer SPIRE bands are likely to be on average at high redshifts.
Some of the bright $500\,\mu$m sources that we detect, especially with a colour selection selecting red galaxies in the sub-mm, may also have their fluxes magnified by lensing and will be among the results forthcoming
from H-ATLAS (Negrello et al. 2010, in prep.).

\section{Conclusions}

We have discussed the spectral energy distribution of sub-mm galaxies in a 14 deg.$^2$, subset of the full H-ATLAS survey.
 We have used the colours in 5 bands to measure the dust temperature
for the sub-sample of sources with known redshifts.
We have also derived a qualitative estimate of the statistical redshift distribution of bright
$350\,\mu$m selected galaxies assuming broad priors in temperature and the dust emissivity.
We found that the average dust temperature of H-ATLAS sources is $28 \pm 8$K, while 
$350\,\mu$m sources with fluxes above 35 mJy   have an average redshift of $2.2 \pm 0.6$.

\begin{acknowledgements}   
Amblard, Barton, Cooray, Leeuw, Serra and Temi acknowledge support from NASA funds for US participants in 
{\it Herschel} through JPL.

Funding for the SDSS and SDSS-II has been provided by the Alfred P. Sloan Foundation, the Participating Institutions, the National Science Foundation, the U.S. Department of Energy, the National Aeronautics and Space Administration, the Japanese Monbukagakusho, the Max Planck Society, and the Higher Education Funding Council for England. 
The SDSS Web Site is http://www.sdss.org/. The SDSS is managed by the Astrophysical Research Consortium for the Participating Institutions.

The UKIDSS project is defined in Lawrence et al (2007).

\end{acknowledgements}

\end{document}